\documentclass[twocolumn]{nature_AP}
\spacing{1.1}

\usepackage{myaasmacros}
\usepackage{graphicx}
\usepackage[top=2cm,bottom=2cm,left=2cm,right=2cm]{geometry}
\usepackage{widetext}
\usepackage[font=small]{caption}
\usepackage{framed}
\usepackage{float}
\floatstyle{boxed}
\newfloat{boxout}{t}{bx}
\floatname{boxout}{Box}

\newcommand*{\vchb}[1]{\begingroup
\setbox0=\hbox{#1}\parbox{\wd0}{\box0}\endgroup}

\def\lsim{\mathrel{\rlap{\lower3pt\hbox{$\sim$}}
    \raise1pt\hbox{$<$}}}                
\def\gsim{\mathrel{\rlap{\lower3pt\hbox{$\sim$}}
    \raise1pt\hbox{$>$}}}                
\newcommand{\kpc}{\mathrm{kpc}}

\newcommand{\Mpc}{\mathrm{Mpc}}
\newcommand{\Msol}{\mathrm{M}_{\odot}}
\newcommand{\pc}{\mathrm{pc}}
\newcommand{\yr}{\mathrm{yr}}
\newcommand{\km}{\mathrm{km}}
\newcommand{\s}{\mathrm{s}}
\newcommand{\cm}{\mathrm{cm}}

\usepackage{color}
\definecolor{fabio}{rgb}{0.3,0.5,0.7}
\definecolor{andrew}{rgb}{1.0,0.2,0.2}

\title{Cold dark matter heats up}
\author{Andrew Pontzen$^{1,2,3}$, Fabio Governato$^4$ \\
$^1$ {Department of Physics and Astronomy, University College London,
  London WC1E 6BT} \\
$^2$ {Oxford Astrophysics, Denys Wilkinson Building, Keble Road,
  Oxford, OX1 3RH} \\
$^3$ {Balliol College, Broad Street, Oxford, OX1 3BJ} \\
$^{4}$ {Astronomy Department, University of Washington, Seattle, WA
  98195, US} \vspace{0.5cm} \\
{\it Draft review as submitted to Nature on 1 Oct 2013.  \vspace{0.2cm}  \\
 Accepted version scheduled for publication on 13 Feb 2014. In accordance with Nature
  policies, the accepted version cannot be posted until 13 Sep 2014.}
\\}

\begin{document}

\begin{widetext}
\maketitle

\begin{abstract}
\noindent

One of the principal discoveries in modern cosmology is that standard
model particles (including baryons, leptons and photons) together
comprise only 5\% of the mass-energy budget of the
Universe\cite{planck}. The remaining 95\% consists of dark energy and
dark matter (DM). Consequently our picture of the universe is known as
$\Lambda$CDM, with $\Lambda$ denoting dark energy and CDM cold dark
matter.  $\Lambda$CDM is being challenged by its apparent inability to
explain the low density of DM measured at the centre of cosmological
systems, ranging from faint dwarf galaxies to massive clusters
containing tens of galaxies the size of the Milky Way. But before
making conclusions one should carefully include the effect of gas and
stars, which were historically seen as merely a passive component
during the assembly of galaxies. We now understand that these can in
fact significantly alter the DM component, through a coupling based on
rapid gravitational potential fluctuations.

\end{abstract}

\vspace{0.5cm}

\end{widetext}

\small

Despite the unknown nature of the dominant components,
$\Lambda$CDM\cite{blumenthal84} successfully describes the evolution
of the Universe from its near-uniform early state, as measured by the
cosmic microwave background\cite{planck}, to the present-day clustered
distribution of matter\cite{percival01} in an accelerating
Universe. 
Consequently the properties of dark matter and the processes driving
the formation and evolution of galaxies are fundamental, closely
connected problems in modern astrophysics.

$\Lambda$CDM, through its explanation of observations on the largest
observable scales, has been established as the standard cosmological
paradigm.  Over time increasingly massive dark matter `halos' form
through gravitational instabilities, starting from small, linear
perturbations in the matter density field.  It is within the
gravitational potential of DM halos that galaxy formation -- gas
cooling and star formation -- proceeds \cite{whiterees78}. However,
long-standing problems have been encountered in reconciling the
predictions of $\Lambda$CDM with observational results at galaxy
scales. These problems likely stem from our poor knowledge of the
complex physics associated with star formation, and are
complicated by failure to identify the DM particle candidate.

The goal of the present review is to present recent progress in
solving the discrepancies. We now understand that gas outflows from
galaxies are ubiquitous, powered by energy released from stars and
black hole accretion. These outflows change the distribution of the
gas and stars which subsequently form. If the outflows launch at
sufficient speed, they also cause an irreversible change in the dark
matter distribution, even if the gas later returns to the galaxy in a
``fountain''. These processes fundamentally modify the structure of
galaxies, and serve to bring theoretical expectations into agreement
with previously problematic observational constraints. It is therefore
important to fully understand the relevant astrophysics before using
galaxies to place constraints on dark matter candidate particles.

\section{Galaxy formation with collisionless cold dark matter}

\begin{figure*}
\includegraphics[width=0.49\textwidth]{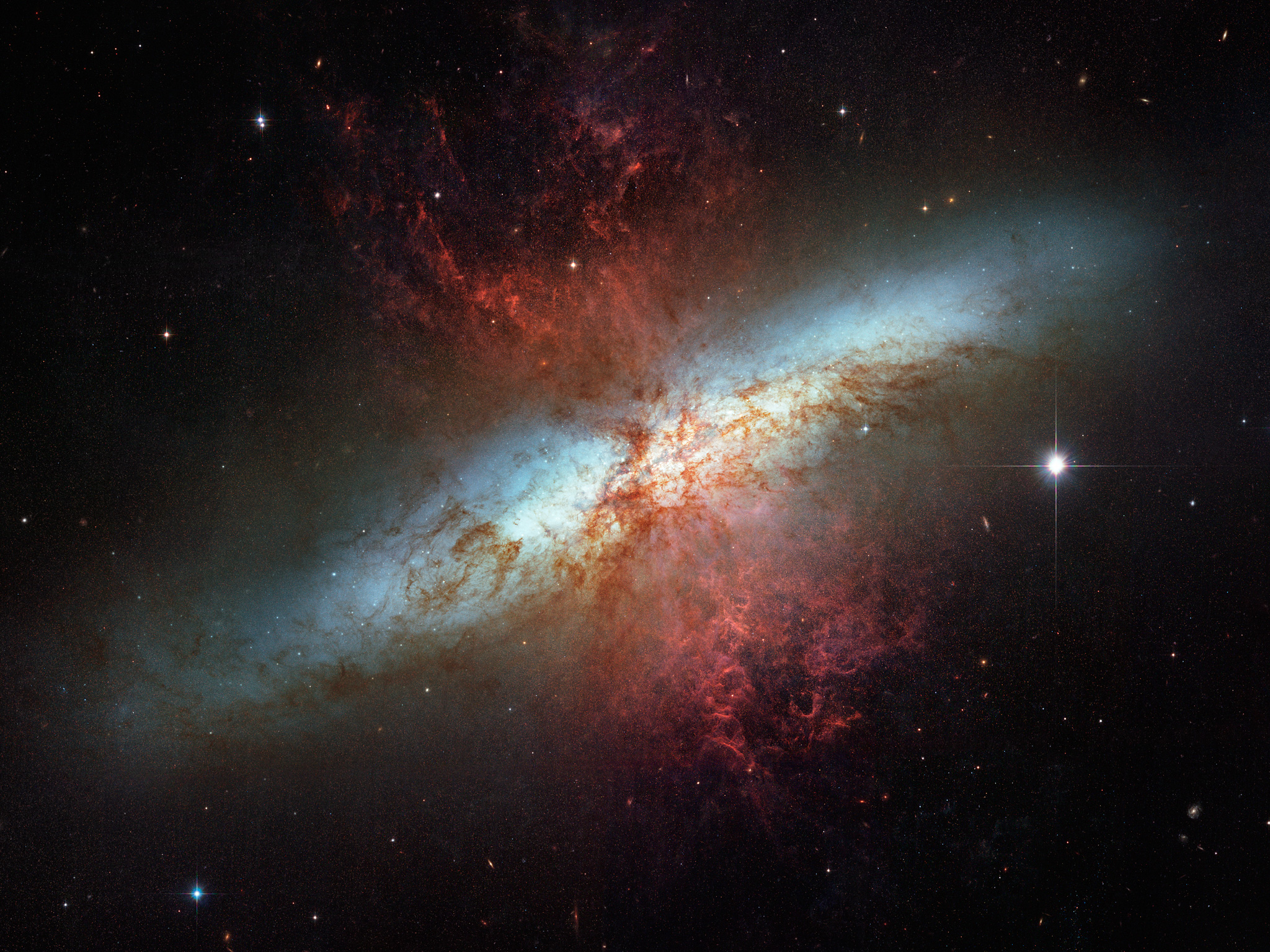} 
\includegraphics[width=0.49\textwidth]{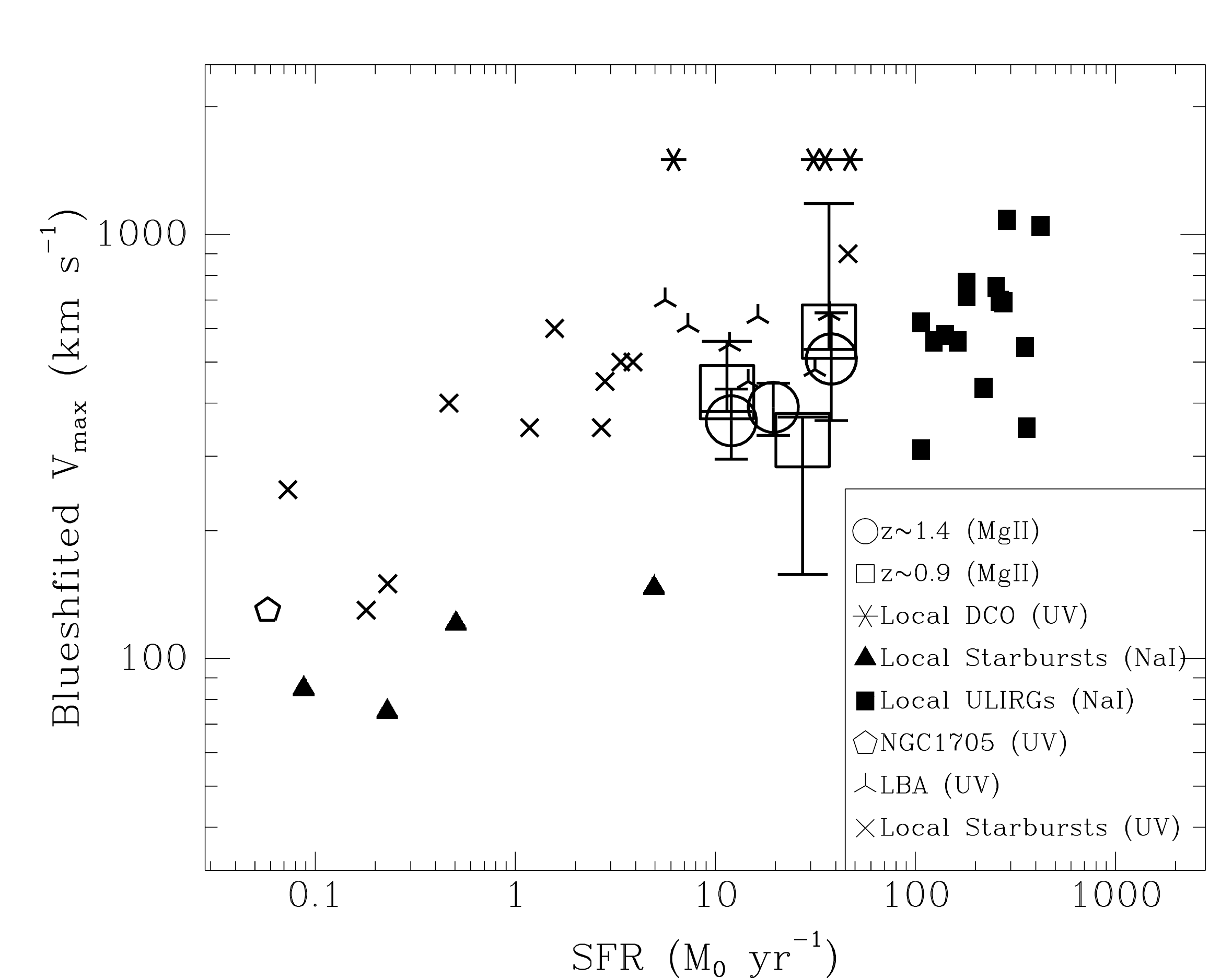}
\caption{The left panel (by J. Gallagher) shows a composite image of
  M82 taken by the Hubble Space Telescope. Purple colours correspond to
  narrow-band H$\alpha$ emission, allowing us to see recombining
  hydrogen in outflowing gas.  The right panel, from Martin et
  al\protect\cite{martin12}, shows a compilation of measured
  absorption line blue-shifts for cool gas as a function of the
  galaxy's star formation rate.
  Even dwarf galaxies with star formation rates
  under $1\,\Msol\,\yr^{-1}$ are able to support winds exceeding
  $100\,\mathrm{km\,s}^{-1}$. The outflow rate of these
  winds is typically several times the instantaneous star formation
  rate of the parent galaxy.}
\label{fig:outflows}
\end{figure*}

The viability of the $\Lambda$CDM picture of structure formation was
first evaluated using computer simulations (allowing, for instance,
neutrinos to be ruled out as the dominant component of dark
matter\cite{frenk85}). Gas cooling and star formation within DM halos
is now the standard paradigm for the origin of
galaxies\cite{whiterees78}. The behaviour of DM can be simulated on
computers by chunking a portion of the universe into ``particles'' and
evolving. Since the particles interact only through gravity, these
simulations are called collisionless.

Early attempts used just $30\,000$ particles to follow large regions
of the Universe.  Consequently one particle had the mass of a large
galaxy -- even so, such simulations were expensive, taking 70 CPU
hours on state-of-the-art 3 MHz facilities. Such calculations would
now take a few minutes on a cellphone.  The growth of computing power
and parallel capabilities meant that, by the 1990s, simulations became
sufficiently powerful to make detailed predictions of the internal
structure of halos in different cosmological scenarios. These
simulations highlighted the universal nature of DM halos formed
through collisionless collapse. The spherically-averaged density of
halos is `cusped' at the centre (scaling approximately as $\rho
\propto r^{-1}$), rolling to a steeper slope at larger radius
(reaching $\rho \propto r^{-3}$); such behaviour is known as ``NFW''
after the authors of a pivotal paper\cite{navarro96}.

At the same time, simulations started highlighting a number of
deficiencies in the CDM scenarios. The most evident was the
overabundance, by more than an order of magnitude, of small
satellites\cite{moore99,klypin99} compared to the number observed
orbiting the Milky Way\cite{mateo98} at the time. Worse, the
simulations significantly over-predicted the density of DM at the
centre of galaxies\cite{moore98}. Increasingly precise observations
of the rotation curves of field galaxies have confirmed this
discrepancy\cite{deblok08} (see \S3).
 
Collisionless DM simulations have since reached maturity, with modern
simulations using several billion resolution elements for just one
Milky Way sized halo\cite{aquarius08,ghalo}.  However to make
predictions which are testable against observations of the real
Universe, baryon physics must be introduced. (Here we are adopting the
astronomical convention of referring to baryons and leptons
collectively as `baryons'.) Because baryons dissipate energy and so
collapse to smaller scales than DM, they constitute a sizeable
fraction of the mass in the central regions of all but the faintest
galaxies \cite{bell01}. Moreover observational constraints on galaxy
formation ultimately come from photons, which can only be sourced by
baryons.  Accordingly much effort has recently been devoted to
implementing gas hydrodynamics and a description of star formation
within simulations\cite{gnedin09,enzo11,ramses,keres12}.

The energy released by young stellar populations and active galactic
nuclei into the surrounding intergalactic medium is critical for
regulating star formation\cite{whiterees78}. Without this energy, most
of the gas becomes cold and dense, rapidly collapsing to form
stars, contradicting observations. Processes providing the energy to halt collapse are
collectively named `feedback' and include supernova winds, radiation
from young stars, and radiation and heat from black hole accretion
\cite{croton06,bower06,Stinson06,hopkins12}. Including these effects has led to
strides forward in forming realistic disk galaxies, reproducing the
efficiency of star formation as a function of galaxy mass, and linking
gas accretion and mergers to galaxy
morphology\cite{robertson06,keres05,dekel09}.  However, until recently
any direct effect of the baryonic component on the DM was limited to a
minor `adiabatic' correction\cite{blumenthal86} (see box A). In other
words, star formation (SF) processes resulted in `passive' changes to the galaxy
population -- modulating the star formation rate without significant
changes to the underlying cosmic DM scaffolding.

This picture has recently been subverted. Spectroscopic observations
reveal the ubiquity of massive galaxy outflows driven by feedback,
carrying significant gas mass away from star forming galaxies
throughout cosmic history\cite{shapley03,weiner09,martin12} (see Section
\ref{sec:evid-galaxy-outfl}).  It has slowly been realised that these
directly observed processes have a non-adiabatic impact on the
associated dark matter halos. The effect is to relieve discrepancies
between baseline CDM simulations and the real Universe (discussed in
Section \ref{sec:evidence-cusp-core}).  The emerging understanding of
these processes constitute the central part of this review (Section
\ref{sec:one-process-rule}).

\begin{figure*}
\vchb{\includegraphics[width=0.49\textwidth]{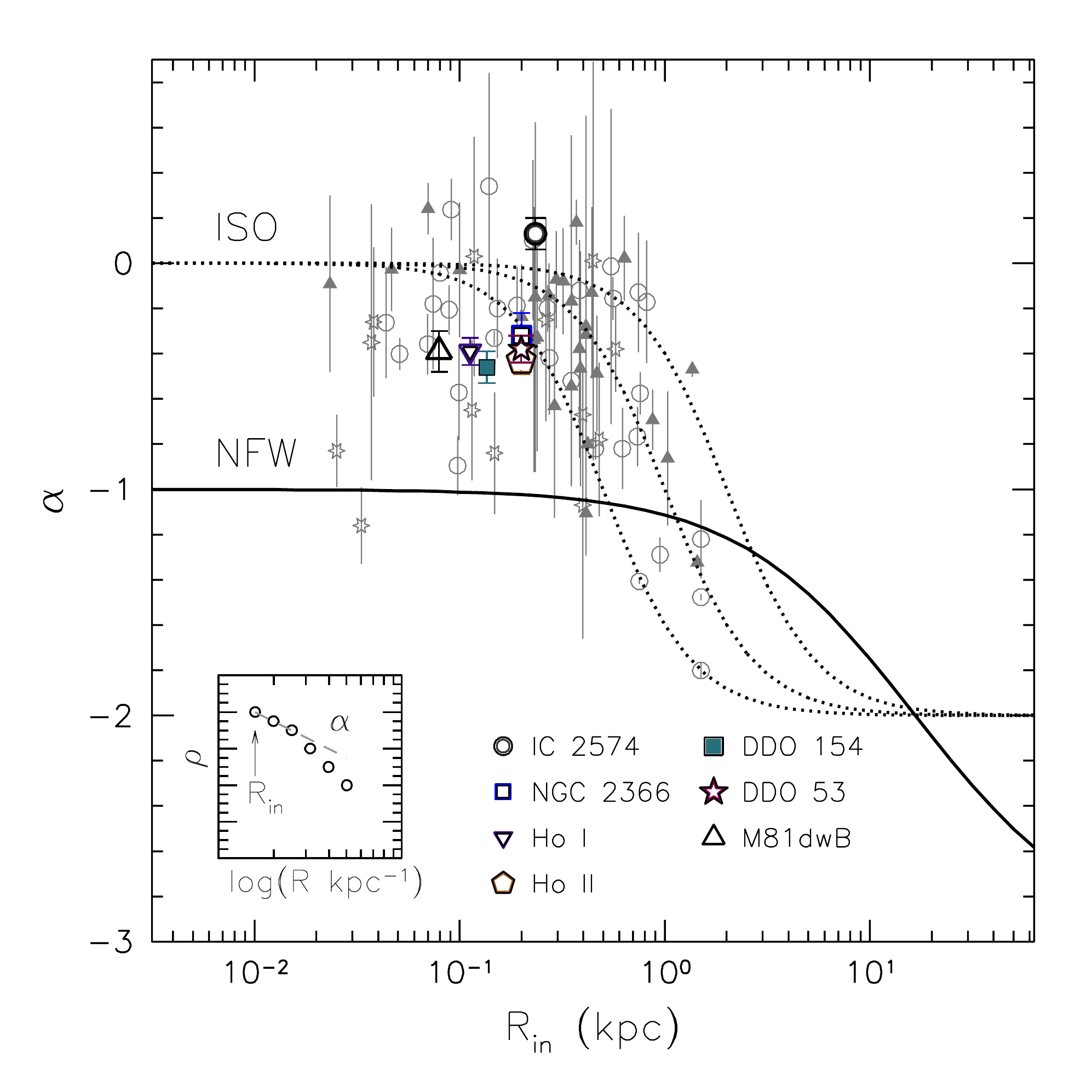}}
\vchb{\includegraphics[width=0.49\textwidth]{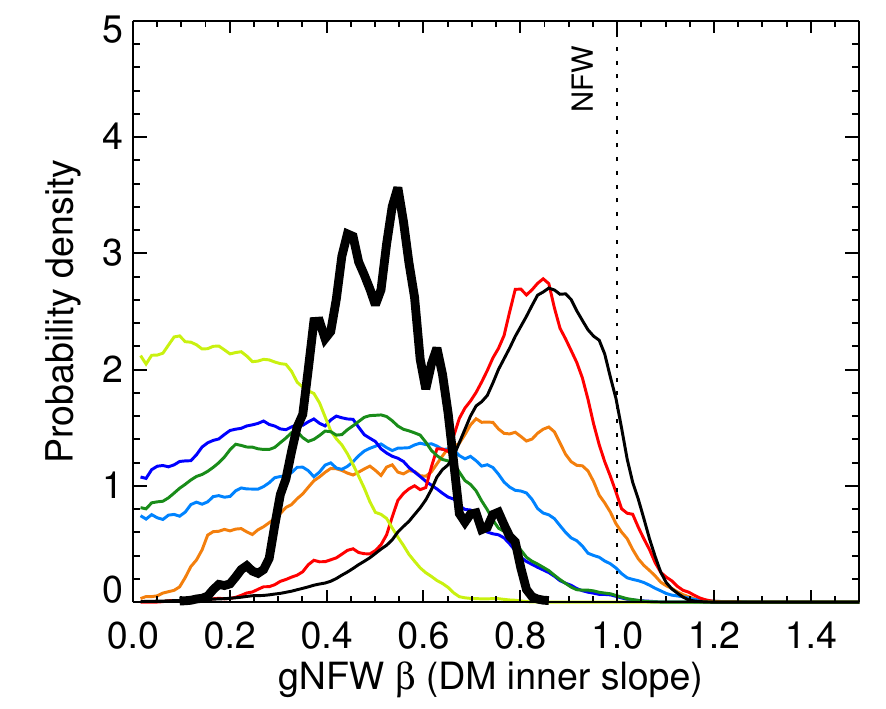}}
\caption{The left panel is a
  compilation\protect\cite{oh11} of observed innermost
  dark matter density profile slopes ($\alpha$ where
  $\rho_{\mathrm{DM}}(r) \propto r^{\alpha}$) for field dwarf
  galaxies, plotted at the innermost point where a robust
  determination has been achieved. Where the slope $\alpha$ can be
  measured interior to around one kiloparsec, it is typically much
  shallower ($\alpha>-1$) than the simulated ``NFW'' result. The right
  panel\protect\cite{newman13} shows the probability
  distribution function on the parameter $\beta=-\alpha$ for a
  selection of galaxy clusters ($M\sim 10^{15}\Msol$). While the
  constraints on individual clusters are quite broad, the combined
  constraints (thick line) again indicate a shallower-than-NFW slope. }\label{fig:cores}
\end{figure*}

\section{Evidence of galaxy outflows and its effect on the stellar component of galaxies}\label{sec:evid-galaxy-outfl}

There is clear observational evidence that star formation activity
drives gas out of galaxies (Figure \ref{fig:outflows}). This largely
arises from studies of the resonance absorption lines imprinted into
spectra by the presence of heavy elements. Consequently dramatic
advances in our knowledge have been made possible by 10m-class
telescope spectroscopy with instruments including Keck
DEIMOS\cite{weiner09} and LRIS\cite{steidel10}. One can either look
for blue-shifted absorption in the spectra of galaxies
themselves\cite{heckman00} or as `intervening' features in the spectra
of background quasars\cite{martin12}. A natural source for the energy
required to generate these outflows is supernovae\cite{maclow99} and
ionising radiation\cite{murray10,wise12} associated with stellar
populations.  In addition, energy released during accretion onto a
massive central black hole may have a role to play, although the
available energy is thought to scale steeply with the black hole's
mass, limiting these effects to the brightest galaxies or their
progenitors\cite{croton06,bower06}.

Recent results underline the ubiquity of outflows\cite{vanderwel11}
and show that their speed likely scales with the star formation rate
of the associated galaxy (see Figure 1). Galaxies are surrounded by
enriched gas moving at hundreds of kilometres per second\cite{rubin10}
in bubbles extending to $100\,\kpc$ or more. This result is
exceptionally hard to explain without significant galactic winds.
Mounting evidence also suggests that much of the in-flowing material
into galaxies may also be metal-enriched\cite{rubin12}, consistent
with a picture in which much of the wind does not attain the escape
velocity but instead re-accretes\cite{shen10,davegang11}.

A separate argument also points to the importance of winds during
galaxy formation. Observed stellar profiles of small galaxies are
mostly `bulgeless', i.e. well approximated by a disk of gas and stars
with an almost exponential profile\cite{dutton09disks,
  kormendy10}. Yet cosmological simulations show that the dark matter
and baryons accumulated in all galaxy halos contain a large fraction
of low angular momentum material\cite{barnes87} -- which would imply
the presence of a bulge\cite{vdbosch01a}. This problem, known as the
`angular momentum catastrophe', is solved if low angular momentum gas
is ejected\cite{binney01,G10} by winds at relatively high-z when SF
peaks\cite{brook11}.  This makes the physics of galactic winds of
fundamental importance to understanding the population of disk
galaxies, even before the effect on DM is considered.

\section{Evidence for a cusp-core
  discrepancy}\label{sec:evidence-cusp-core}

We now turn our attention to the excessive quantity of dark matter
predicted by the CDM model compared to measured densities in the
innermost regions of galaxies and clusters.

\subsection{Dwarf galaxies}

As explained above, the under-abundance of dark matter in the centre
of dwarf galaxies relative to theoretical predictions is known as the
cusp-core discrepancy. The problem was discovered as soon as
cosmological simulations became capable of predicting halo
structure\cite{flores94,1994Natur.370..629M}. Although acceptance was
gradual, it is now firmly established that robust measurements of the
dark matter density can be made from rotation curves of gas-rich dwarf
galaxies `in the field' (i.e. away from the influence of larger
galaxies). In the innermost regions $r\lsim 0.5\,\kpc$ the baryonic
contribution to the potential is comparable to that of the dark matter
and must be subtracted\cite{blitz05,deblok08}. Consequently inferring
the dark matter density requires (1) high spatial resolution of the
gas and stellar kinematics (2) a comprehensive understanding of how to
estimate and subtract the stellar and gas mass distribution from the
central kiloparsec and (3) careful handling of systematic
observational errors.  The last category encompasses possible biases
arising from radio beam-smearing, departure from circular orbits,
centring difficulties, unknown details of stellar mass-to-light ratios
and gravitational potential asphericity within galaxies; these are now
thought to be under control, since we can test algorithms on mock
observations from simulations (where the true density is
known)\cite{swaters03,oh11,oh11sim}.
 
Results from recent surveys of the local Universe such as THINGS and
LITTLE THINGS \cite{walter08,hunter12} can therefore be regarded as
free from significant observational bias.  These samples reveal
shallower-than-NFW dark matter profiles in a large fraction of dwarf
field galaxies, with $\rho \propto r^{-0.4}$ interior to $r\simeq 1
\,\kpc$ (Figure \ref{fig:cores}, left panel). The objects are referred
to as `cored' although the estimated density profile is almost never
actually flat. After 20 years of study the cusp-core problem has
remained a persistent and significant discrepancy between theoretical
models of a $\Lambda$CDM universe and observations of dwarf galaxies.

\subsection{Milky Way Satellites}

Small galaxies known as `dwarf spheroidals' orbit close to the Milky
Way. The dwarf spheroidals have little gas content and their stellar
content is not in a rotational disk\cite{mateo98}. This likely
reflects the effect of tidal fields and strong interactions with the
hot gas in the halo of the parent galaxy\cite{mayer01}.  Sampling the
smallest halo masses in which galaxies form, these satellites have the
potential to constrain the properties of dark matter and the physics
of galaxy formation and have accordingly received significant
attention\cite{nierenberg12}.

We discussed above how field dwarfs have been fundamental in revealing
the apparent over-concentration of DM at the centre of
halos. Satellite dwarfs, with an order of magnitude fewer stars still,
are potentially powerful probes of the DM distribution at the smallest
scales\cite{strigari08}.  Various techniques hint at the existence of
cores, rather than cusps, in the brightest dwarf
spheroidals\cite{goerdt06,walker11,breddels13}. However because galaxy
satellites do not possess HI disks and deviate from spherical
symmetry, inferring the mass distribution of their DM halos is
significantly harder than for field galaxies.
Simpler is to measure total mass inside the half-light radius (which
typically lies at a few hundred parsecs)\cite{wolf10}. Compared to the
most massive satellites in CDM it is widely believed that there is too
little mass in each real dwarf spheroidal, a problem which is referred to
as the objects being ``too big to fail'' \cite{bk11b}. However the
effect of tidal forces and stripping\cite{zolotov12,arraki12}
complicate the interpretation.  At present the properties and
abundance of isolated, small field galaxies provide stronger
constraints on models of SF and feedback and alternative DM
models\cite{papastergis11,ferrero12}.

\subsection{High mass galaxies and galaxy clusters}

Field dwarfs typically fall into the category of
``low-surface-brightness'' galaxies, defined by their extended diffuse
stellar and gaseous disks. The uncertainties (discussed above) in
recovering dark matter distributions in these objects are mitigated by
the relatively small baryonic contribution to the potential at the
time they are observed. A fraction of more massive galaxies (with
rotational peak velocities larger than $100\,\km\,\s^{-1}$) also have
these characteristics.  Analysis of such galaxies\cite{kuziodenaray11}
again point to relatively flat central DM profiles. This is a
significant finding because it shows that cores can be formed in halos
with estimated stellar masses up to $5 \times 10^9 \,\Msol$.

The inner distribution of DM in galaxies with more conventional,
massive disks (similar to our own Milky Way, for instance) is
unfortunately harder to ascertain because the gravitational potential
is more strongly dominated by baryons\cite{bell01}, so that
uncertainties in the age, metallicity and hence light-to-mass
conversion ratios of stellar populations dominate. However, many
attempts have pointed to smaller central dark matter densities than
theoretically expected\cite{mcgaugh07}, in line with the
low-surface-brightness results. Some observations point to well
defined scaling laws that link the DM and baryon components, with DM
and baryons following similar profiles\cite{gentile09}.  The
significance of this relation is still very poorly understood but it
may point to a tight coupling between baryons and DM at galactic
scales\cite{swaters11}.  More indirect constraints on the central DM
densities in luminous galaxies arise from the existence of stellar
bars\cite{perez12} which, over cosmological timescales, seem
dynamically incompatible with the presence of cuspy dark matter
halos\cite{debattista98}.

The largest bound systems in the Universe, galaxy clusters, have a
mass $\gsim 10^{14}\,\Msol$, comparable to a hundred or more Milky Way
galaxies. The dark matter distribution in these objects can now be
measured by a number of independent techniques, making them one of the
most interesting cosmic laboratories to study baryon and DM
interactions.  Their central density is sufficiently high that strong
gravitational lensing\cite{miralda95} constrains the mass on scales of
$\sim 10$ to $100\,\kpc$; statistical weak gravitational
lensing\cite{umetsu11} can be applied on scales between $100\,\kpc$
and a few $\Mpc$; and information from the kinematics of member
galaxies and the brightest cluster galaxy (BCG) stars\cite{lokas06} or
gas distribution\cite{allen01}
gives constraints at scales of below $\sim 10\,\kpc$.  Taken together, these
multi-wavelength observations provide tracers of the total density
profile over multiple scales of interest, from a few kiloparsecs
outwards.  Current state-of-the-art studies that combine the above
approaches have recovered a total density profile that is essentially
compatible with NFW at all scales; however, once the stars of the BCG
are subtracted, the dark matter has a central profile shallower than
NFW\cite{newman13} on scales of $\sim 10\kpc$. A wide range of
possible explanations for these indications of a `universal profile'
not for DM, but rather for collisionless matter (comprising stars and
DM together) have been proposed. We will explore these in Section
\ref{sec:one-process-rule}.

\section{Gravitational interactions between baryons and DM}\label{sec:one-process-rule}

\begin{boxout*}[t!]
\small 
{\bf Box A --- how gas affects dark matter through gravity }
\begin{center}
\includegraphics[width=0.8\textwidth]{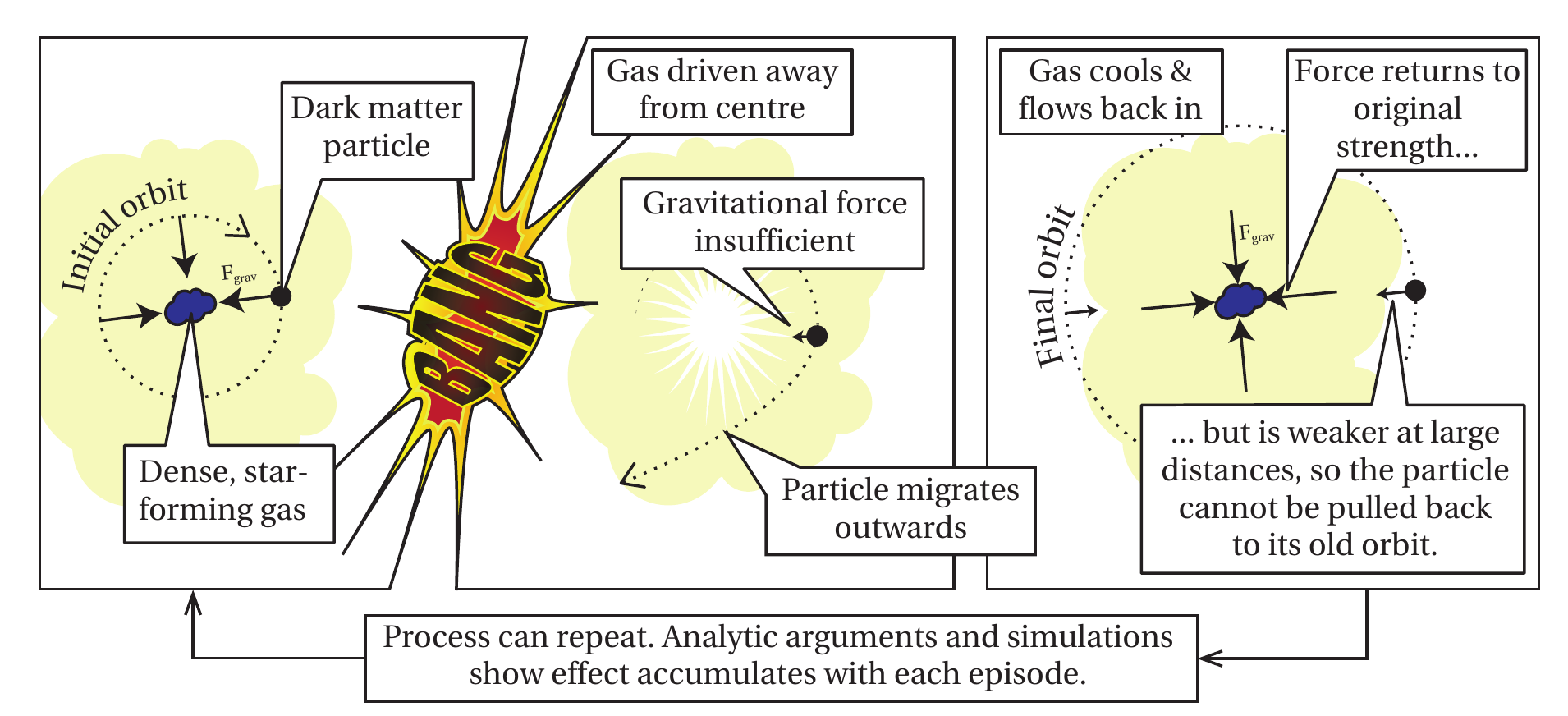}
\end{center}

\setlength{\parskip}{5pt}
\setlength{\parindent}{5pt}

The simplest known mechanism with which baryons and DM exchange energy
through gravity is called {\bf adiabatic
  contraction}\cite{blumenthal86}. The word `adiabatic' refers
to a slow deepening of the gravitational potential as gas
gradually accumulates in the centre of a dark matter halo on
timescales longer than the local dynamical timescale.  The added
gravitational attraction of the accumulated material causes the dark
matter to contract.

If gas arrives not in a smooth flow but in dense, discrete chunks
(i.e. infalling satellite protogalaxies), this picture may be
qualitatively modified by \textbf{dynamical friction}\cite{elzant01}.
This effect is usually pictured as a gravitationally-induced density
wake behind infalling dense clumps -- the wake pulls back on the clump
with the result that the kinetic energy of the clump is transferred
into the dark matter.

The assumptions underlying adiabatic modelling can also fail due to
\textbf{outflows} if these evacuate gas at speeds significantly
exceeding the local circular velocity\cite{navarro96cores}. Under the
adiabatic approximation, removing gas would be expected to simply
reverse the effects of accumulating it in the first place, so that the
final energy of any given dark matter particle would be
unchanged\cite{gnedin02}. However if the removal proceeds sufficiently
quickly, net energy is transferred into the dark
matter\cite{read05,PG12}. Moreover this transfer is irreversible in
the sense that re-accreting the lost gas does not lead to a
compensating energy loss\cite{PG12} (see box figure). These results hold even if the gas never
leaves the galaxy but is simply moved in bulk
internally\cite{weinberg02,mashchenko06,PG12}.

The reason for this is as follows\cite{PG12}. Consider a dark matter
particle that orbits close to the centre of the halo, where the gas is
dense. If the gas is locally removed on a short timescale, the
gravitational centripetal force holding the dark matter in its orbit
instantaneously vanishes (or, rather, is substantially reduced in
magnitude). The dark matter particle responds by flying outwards. Even
if the gas later returns, the dark matter particle resides further
away from the centre by the time this reversal occurs. The $1/r^2$ law
of gravity means the increase in force felt by the particle is quite
small compared to the force originally holding the particle near the
centre. The particle therefore continues to live at a large radius;
this implies a net gain in energy. Repeating the process has an accumulative effect, which
allows a significant transformation to be accomplished by recycling a
small amount of gas instead of expelling an unfeasibly large amount of
gas in one episode.

\end{boxout*}

We have outlined above the observational evidence pointing towards
systematic departures of the distribution of dark matter from the
original expectations of the CDM paradigm.  It has been widely suggested
that this discrepancy could be addressed by gravitational interactions
(the only way baryons and CDM can interact) that transfer energy from
the baryon component to the diffuse dark
matter\cite{navarro96cores}.
If sufficient energy can be given gravitationally to dark matter
particles in the centre of the halo, they will then migrate outwards,
reducing the central density (note that this process will also apply
to the stellar
component\cite{2005MNRAS.356..107R,PG12,teyssier13}). Energy can be
transferred between these two components in two distinct ways: from
the kinetic energy of incoming material or from baryonic processes
linked to feedback within the galaxy. We will tackle these
possibilities in turn.

As dense clumps move through a diffuse DM background a fraction of the
orbital energy of the incoming material is lost to internal energy of
the diffuse halo through ``dynamical
friction''\cite{white76,elzant01,Tonini2006,romano08} (see Box A for an
explanation). The sinking of dense gaseous or stellar clumps can 
flatten the central DM profile over a range of scales, although
significant core creation has only been demonstrated in simulations
of galaxy clusters rather than at the scale of individual
galaxies\cite{elzant04}. Note that dense, centrally-concentrated
baryons in in-falling clumps are an essential pre-requisite in this
process.

The second class of energy sources comes from within the galaxy
itself: energy liberated from stellar populations can be large
compared to the binding energy of the galaxies\cite{binney01}. Early
work suggested that removing most of the baryons in a rapid, dramatic
starburst event could over-compensate for the previous adiabatic
contraction, leading to the desired effect of reducing the central DM
density \cite{navarro96cores}.  Subsequent works studied the
feasibility of this mechanism in more
detail\cite{gnedin02,mo04,read05}, showing in particular that repeated
outflow episodes interspersed by reaccretion had a cumulative effect
on the dark matter\cite{read05}. 

However these early investigations were limited by the unknown
behaviour of gas in dwarf galaxies over cosmic time, and the lack of
any clear analytic framework for understanding the apparently
irreversible response of the dark matter. It was unclear even to what
extent the available energy in stellar populations couples to the gas
through heating and radiation pressure; consequently the idea of
energy transfer from baryons to the DM was not widely accepted at this
stage.

Other authors\cite{weinberg02,mashchenko06} showed that gas remaining
fully within the system can still be effective in removing cusps when
coupled to an energy source such as stellar feedback. For instance
supernovae driving gas on timescales close to the local orbital period
was identified as a mechanism to transfer energy to dark matter
particles\cite{mashchenko06}. In this case the cusps were destroyed in
an energetically consistent manner without requiring any
unrealistically dramatic outflows.
By 2008 advances in numerical resolution and understanding of how gas
cools before forming stars allowed for realistic treatments of the
relevant hydrodynamics (Box B). Simulations at high
redshift\cite{mashchenko08} showed that dark matter could indeed be
expelled self-consistently from the central regions of small
protogalactic objects.  This work provided the first proof-of-concept
in a cosmological setting, but did not make predictions of observable
objects (dwarfs, for being faint, are only observable in the nearby,
redshift-zero Universe).

As it became possible to resolve star forming regions\cite{saitoh08}
throughout the assembly of a dwarf galaxy from the young universe to
the present day, for the first time simulations formed galaxies with
stellar, gas and dark matter distributions consistent with
observational bounds\cite{G10,G12,munshi12}. Multiple short, locally
concentrated bursts of star formation were the key new phenomenon
enabling modification of the DM distribution: by temporarily
evacuating gas from the central kiloparsec of the galaxy these cause
dark matter to migrate irreversibly outwards\cite{PG12}; see Box
A. The actual process in play thus combines characteristics of the
multiple-epoch outflow picture\cite{read05} and the internal-motions
picture\cite{mashchenko06}. It does not require fine-tuning of the gas
velocity or dramatic evacuation of the gas from anything but the
innermost region.  The key requirement is that the gas exit the centre
of the galaxy faster than the local circular velocity.

Analytic modelling of multiple, impulsive changes to the gravitational
potential gives considerable insight into how these changes arise and
why they are irreversible\cite{PG12}.  This allows for an accumulation
of effects as the process repeats in several gas outflow events. In a
single event the total gas mass in the galaxy limits the effect of
outflows\cite{gnedin02} but when the same gas is recycled and used in
multiple events the only practical limitation is the total energy
liberated from stellar populations and black holes (see below).  
The model of core creation through repeated outflows draws strong
support from both analytic arguments\cite{PG12}, and simulations using
different numerical techniques\cite{teyssier13}.  Observationally,
dwarf galaxies, where the evidence for cores is strongest, are
observed to be gas rich and show evidence for repeated small bursts
and prolonged star formation histories\cite{mcquinn10}.  This supports
a picture where the effect on the dark matter builds up over several
Gyrs\cite{read05,PG12}, during which gas is being cycled in repeating
outflow and cooling episodes.

\subsection{Scaling with mass and the significance of satellite galaxies}\label{sec:scaling-with-mass}

\begin{figure}[t]
\begin{center}
\includegraphics[width=0.5\textwidth]{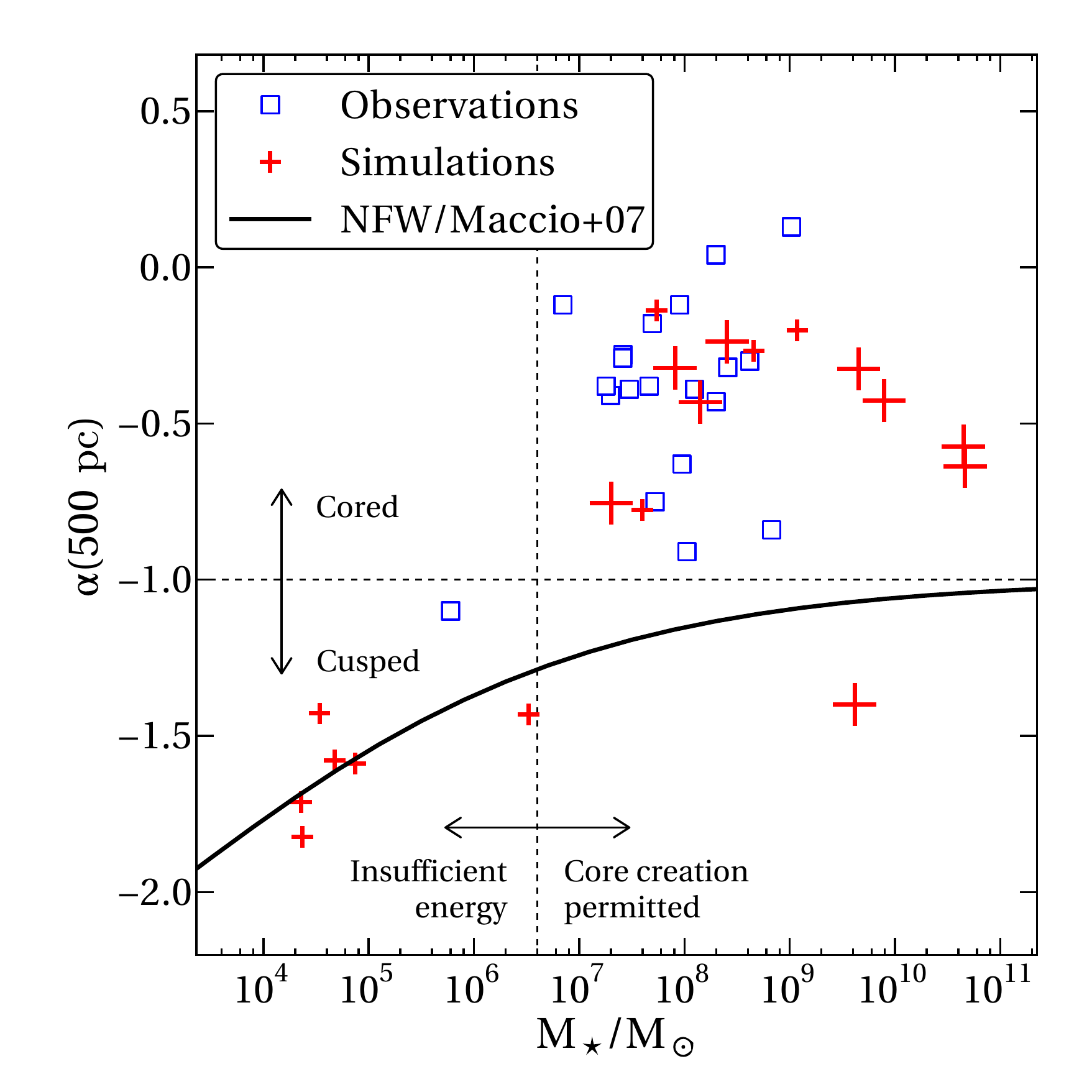}
\end{center}
\caption{The log slope of the dark matter density (at radius
  $500\,\pc$) plotted against mass of stars formed (updated from
  Governato et al\cite{G12}). The expected slopes from pure dark
  matter calculations are approximated by the solid line, while
  hydrodynamic simulations have shallower slopes indicated by the
  crosses. When less than $\sim 10^{6.5}\,\Msol$ of gas has formed into
  stars, there is insufficient energy available to flatten the
  cusp\cite{penarrubia12}. The boxes show data from the THINGS
  survey\cite{oh11} of field dwarf galaxies. Additional observational
  data at stellar masses lower than $10^6\,\Msol$ would be highly valuable. }
\label{fig:when-cusps-flatten}
\end{figure}

\begin{boxout}
\small
{\bf Box B --- why high resolution gas dynamics generates outflows}

\setlength{\parskip}{5pt}
\setlength{\parindent}{5pt}

Computer simulations of the formation of galaxies would ideally
resolve cosmological large scale structure (on 10's of megaparsecs)
down to the scale of individual stars (at least $10^{14}$ times
smaller). This is, and seems certain to remain, unfeasible.

The approach is instead to mimic the effects of stars without actually
resolving them individually. Since star formation is the
conclusion of run-away gas cooling and collapse, a typical
computational approach is to form stars when gas satisfies certain
averaged conditions, and in particular when it reaches a
certain threshold density. But as resolution slowly improves in
simulations , smaller regions and larger densities can be
self-consistently resolved\cite{saitoh08}.

Until the mid-2000s, a typical threshold density was set at
$0.1\,m_H\,\cm^{-3}$, where $m_H$ is the mass of a hydrogen atom. This
corresponds to the mean density of galactic neutral atomic gas, so
stars form throughout the disc of a typical simulated galaxy. Energy
output from stars in the diffuse medium results in a gentle heating of
the entire galaxy, slowing the process of further star
formation. 

However if one can achieve sufficient resolution (and implement the
more complicated cooling physics
required\cite{gnedin09,shen10,christensen12a}) to push to $10$ or
$100\,m_H\,\cm^{-3}$ qualitatively different behaviour results. This
is the density that corresponds to molecular clouds in our galaxy,
known to be the sites where clusters of stars form. Instead of forming
stars in a diffuse way through the entire disc, one now efficiently
forms stars in small, isolated regions\cite{G10,hopkins12}, which is
considerably more realistic.

When energy from the resulting stellar populations is dumped into the
gas, it heats to much higher temperature than diffuse star formation
achieves. It is likely that intense radiation
pressure is also a significant factor\cite{murray10}. In any case, the
gas is over-pressurised by a factor of at least $\sim 100$ compared to
its surroundings and expands rapidly. 
The combination of high initial density and explosive decompression is
suitable for launching galactic-scale outflows; but it is also what
allows an efficient coupling of the available energy to dark matter
(box A).

\end{boxout}

A key part of confirming which mechanisms are responsible for flattened
dark matter profiles is to predict and understand in detail how the
processes affect systems of differing mass.
Building on the impulsive picture\cite{PG12}, full numerical
simulations\cite{G12} and analytic arguments\cite{penarrubia12} 
have all pointed to a transition between core creation and persistent
cusps below a critical stellar mass. This dividing line likely lies
between $10^6$ and $10^7\,\Msol$ (assuming most of the energy
available from supernovae is transferred to the dark matter). For less
massive stellar systems, the direct effects of stellar feedback on the
dark matter should be minor on energetic grounds
alone\cite{penarrubia12}, as SF becomes less efficient; see Figure
\ref{fig:when-cusps-flatten}. The energetic argument shows that the
possible cores from supernova feedback would be indetectably small for
stellar masses significantly below $10^6\,\Msol$.

For stellar masses exceeding $10^7\,\Msol$, it is clear that energy
from SF processes is available to alter the central regions of the
dark matter halo through sufficiently rapid galactic fountains or
outflows\cite{G12}, but few simulations of luminous galaxies reach the
resolution necessary to study the formation of cores.  The Eris
simulation (a high resolution simulation of a Milky-Way analogue) has
recently been reported\cite{kuhlen13} to have a dark matter core on
scales of around $1\,\kpc$. On the other hand it has been reported
that cores shrink with respect to the halo scale
radius\cite{dicintio13} for masses exceeding $10^{11}\,\Msol$ (the
Milky Way mass is $\sim 10^{12}\,\Msol$). These statements may be
reconcilable; further higher resolution work is required for progress
in our understanding. As masses continue to increase to the cluster
scale (see Section \ref{sec:evidence-cusp-core}), further processes
become interesting. For instance numerical work has shown that
accretion onto the central black hole, if proceeding in repeated,
highly energetic bursts, replicates the effect of supernovae on dwarf
galaxies\cite{martizzi12}.

\subsection{The alternative: modified dark matter}

Many possible processes which can change the dark matter distribution
in the centre of galaxies assume that the dark matter particle is cold
and collisionless (i.e. interacts only through gravity) -- a `minimal'
scenario. However the observational controversies detailed in Section
\ref{sec:evidence-cusp-core} have prompted considerable interest in
non-minimal DM models. By changing the properties of the dark matter
candidate particle, the predictions for the distribution within halos
is altered; potentially, therefore, galaxies and galaxy clusters
become an important probe of particle physics\cite{markevitch04}.  For
instance, the class of warm dark matter\cite{boyarsky09} models (WDM)
invoke a candidate particle with non negligible residual streaming
motions after decoupling (such as a sterile neutrino), suppressing the
formation of small scale structure\cite{dalcanton01} and delaying the
collapse of dwarf sized halos and their associated star formation to
slightly later epochs \cite{menci12}. On the other hand these models
do not produce cores on observationally relevant
scales\cite{kuziodenaray11} and are currently strongly constrained by
the clustering of the neutral gas in the cosmic web\cite{viel13}.
Self-interacting dark matter (SIDM)\cite{spergel00}, on the other
hand, refers to particle physics scenarios with significant 'dark
sector' interactions. SIDM behaves more like a collisional fluid,
preventing the central high-density cusp from forming and makes the
central regions more spherical\cite{peter12shapes}. Unlike in the WDM
case, the number density of DM halos remains relatively unchanged even
at the smallest scales\cite{zavala13}.  The diversity of theoretical
models, however, gives significant freedom in the choice of the cross
section and its possible dependence on particle velocity
\cite{zurek12}.  This makes it difficult to establish a single
baseline SIDM scenario.

Overall it seems that neither WDM nor SIDM on their own provide a
complete alleviation of the tensions detailed in Section
\ref{sec:evidence-cusp-core}.  In particular, because the infall
pattern of matter is driven by the large structure, no DM model can
alone alleviate the problem of removing low angular momentum baryons
from the centre of galaxies without unfeasible modifications to the
large scale power spectrum of matter fluctuations. But the effects of
baryons may amplify or change the signatures of these particle models
(or, worse, make them more similar to the prediction of the CDM
model).  The dwarf spheroidals teach us that different transformative
mechanisms interact in surprising, non-linear ways\cite{zolotov12},
motivating a more detailed study of the galaxies formed in fully
hydrodynamical simulations with WDM or SIDM.

Ideally to alleviate degeneracies between particle-physics and
outflow-induced modifications to CDM, one would identify regimes in
which only one or the other is active. This points towards the future
value of careful studies probing scalings of
cores from stellar masses below $10^7\,\Msol$
(where the energy available to create cores is so limited that
baryonic effects are tightly constrained) to above $10^{13}\,\Msol$
(where a variety of processes are feasible).

\section{Conclusions}

The $\Lambda$CDM cosmology underlies a highly successful paradigm for
explaining the formation of visible structure in the universe. Until
recently, the key ingredients were passive processes which controlled
the association of observable matter with the dark matter (for
instance suppressing over-efficient star formation) while having
little explicit effect on the underlying dark matter. There is,
however, a new, rich literature of processes which violate this basic
assumption and lead to fundamental modifications to the observable
properties of galaxies. In the last few years these have come into
sharp focus as increasingly sophisticated computer simulations have
begun to follow the effects of star formation, and many relevant
observational techniques have matured to the point that they can be
regarded as robust.  Direct evidence of precisely which `baryonic
processes' are in play and their relative importance in the real
Universe at different scales should be our next priority. Because
these baryonic processes simultaneously modify a number of
observational diagnostics (outflows, dark matter cores, stellar
morphology and star formation regulation), they weave into a coherent,
testable framework.

It remains a possibility that tensions between observation and theory
at the scale of faint dwarfs and clusters may point to exotic particle
physics.  Ultimately we expect that a concerted effort from theorists
and observers can achieve the goal of pointing to unique predictions
of non-minimal DM models.  Of particular interest in the coming years
will be ({\it i}) improved understanding of the dark matter in dwarf
spheroidals and faint field galaxies; if cores persist at the faintest
end, it is a generic conclusion that baryonic physics cannot account
for them\cite{G12,penarrubia12}; ({\it ii}) study of the stellar
population ages and, separately, metallicity distributions of these
objects to determine as far as possible whether the required bursty
star formation histories are consistent propositions\cite{mcquinn10};
({\it iii}) better predictions of the scalings of cores in massive
galaxies and clusters for different scenarios; ({\it iv}) observations
that constrain the star formation histories of dwarfs\cite{pacucci13}
and the behaviour of gas at high redshift, especially through
absorption line studies which are sensitive to internal kinematics and
outflows\cite{viel13}; ({\it v}) renewed effort to understand how
non-minimal dark matter scenarios (such as WDM or SIDM) interact with
the revised, more complex baryonic physics of galaxy formation.

\subsection{Acknowledgements}

We would like to thank Se-Heon Oh, Simon White, Max Pettini, Crystal
Martin, Matt Walker, Jorge Pe{\~n}arrubia, Alyson Brooks, Tommaso
Treu, Richard Ellis, James Wadsley and Lisa Randall for helpful
discussions and comments on an early draft.

\section*{References}

\footnotesize
\vspace{1cm}
\bibliographystyle{naturemag}
\bibliography{nature_review.AP_submit.bbl}

\clearpage

\end{document}